# An astronomical survey conducted in Belgium

Yaël Nazé[1], Sébastien Fontaine[2]

[1] *Research associate FNRS, GAPHE, Department AGO, University of Liège, Belgium*

[2] *Social & Human Sciences Institute, CLEO, University of Liège, Belgium*

Abstract: This article presents the results of the first survey conducted in Belgium about the interest and knowledge in astronomy. Two samples were studied, the public at large (667 questionnaires) and students (2589 questionnaires), but the results are generally similar in both samples. We evaluated people's interest, main information source, and attitudes towards astronomy, as well as their supposed and actual knowledge of the subject. The main conclusion is that, despite a poor self-confidence, people do know the basic astronomical concepts. However, that knowledge is not deeply rooted, as reasoning questions show well-spread misconceptions and/or misunderstandings.

## 1. Introduction

Improving the science literacy level is not an easy enterprise, but it is certainly doomed to fail if the actual whereabouts of the public are unknown as several studies showed (e.g. Danaia & Mc Kinnon, AER, 6, 32, 2008, and references therein). Assessing the knowledge and appreciation of the public is thus a prerequisite before an efficient science diffusion endeavor can be undertaken.

Our aim was to meet that goal for a specific field, astronomy and space, and in a specific region, the French-speaking, southern part of Belgium. In that zone, there are a number of amateur societies, highly interested and well informed in the subject, but they represent only a small fraction of persons. To give an example, the largest amateur society in the surveyed area is the *Société Astronomique de Liège* that counts 700 members. These amateurs correspond to the most "attentive" category of public, following the criteria of Miller (1983, The American people and Science policy, NY:Pergamon), but they are certainly not representative of the majority of the public. We repeatedly experienced this contrast in our outreach activities and felt that most public talks and press releases were obviously too technical and/or too far from people's interest. We therefore undertook this survey, with the goal of pinpointing people's interest, assessing their astronomical literacy level and misconceptions, and improving our outreach practices.

While many such astronomical surveys have been performed in the USA (e.g. numerous articles in the astronomy education review), none had ever been conducted in Belgium. Only the general Eurostat barometers about citizens and science and the general PISA science barometers of students are available. As the Belgian scores were in these cases generally close to the European averages, it shows the potential of the country as a reference. However, the Eurostat barometers are general, with only three items related to astronomy and space. The first one is a question about the most interesting science & technology developments: about 20% of the public chose the item "astronomy and space" (17% in the 2001 barometer, 23% in the 2005 barometer), which puts this subject amongst the ones arising the least interest (with genetics and nanotechnologies). This result depended on gender (men choosing this item twice more often than women), age (young more interested than old) and education (positive correlation with diploma level). A similar result was obtained in the 2008 barometer examining young (15-25 year old) people: 22% show a high interest in news related to "the universe, the sky and stars", 42% being moderately interested and 36% not at all interested. The second item asked people to judge whether a discipline was scientific or not on a scale between 1 (not at all scientific) to 5 (very scientific): in 2005, 70% of people considered astronomy as very scientific or scientific (choice of 4 or 5 on the scale), while 41% did the same for astrology (and this decreased to only 13% when the word "horoscope" was used). Astronomy thus appeared amongst the disciplines regarded as highly scientific, together with medicine, physics, biology, and mathematics. The third occurrence appears in a series of true/false statements aimed at assessing the scientific literacy level where two items in the 2005 survey were linked to astronomy (Sun going around the Earth and Earth revolving around the Sun in one month): 66% of Europeans chose the correct answer.

We wanted a more complete picture than these three items in the European studies could provide, and therefore designed a dedicated astronomical survey. It is presented in the next section, before presenting the results obtained.

## 2. Design of the study and methodology

2.a. Design of the questionnaire

The questionnaires consist of 4 parts. The first section contains the socio-demographic factors: gender, age, education level (current for the students, achieved for the others) and occupation. They are indeed well-known factors influencing the results (e.g. Eurostat barometers, Entradas et al., PUS, 22, 269, 2013 and references therein). To complement these usual criteria, we added questions on religious beliefs and interest in ecology. There are also questions enabling us to identify specific groups which could potentially bias the results, i.e., people having a scientific formation, having followed astronomy courses, or being amateur astronomers. Finally, people were asked to self-evaluate their cultural level and their astronomical level.

The second part of the questionnaire is a series of questions assessing the interest for astronomy matters, as well as simple practices (search for information on the subject, discussion about the subject, type of performed sky observation).

The remaining of the questionnaire is related to knowledge and understanding of astronomical concepts. This part is inspired by the ADT (Astronomy Diagnostic Test, Hufnagel, AER, 1, 47, 2002), but not identical to it, for several reasons. First, astronomy is not part of the Belgian curricula of primary or secondary schools: we thus cannot test whether taught concepts were assimilated. In Belgium, astronomical teaching exists only in two forms: dedicated university courses containing astronomy, which are reserved to bachelor and/or master degrees in physics-astronomy, and astronomy courses organized by some amateur societies for their members. The first part of the questionnaire enables us to identify the few people in this situation. Second, we decided to focus on astronomy, avoiding pure physics: no questions about gravity or weight were asked. Third, adding figures to the questionnaires was not possible for practical reasons, which again eliminated some types of questions. Finally, we decided to avoid jargon (such as quasar, pulsar, or CMB) and complex concepts (link color-temperature, nature of light,…), to stick as much as possible to everyday experience.

We otherwise kept the spirit of the ADT. For example, we explored only one concept per question. In our questionnaire, most of the questions were true/false statements, with only a few open questions (when an estimate of size or distance is asked) and a few multiple-choice questions. In the latter case, we made sure that the answers were mutually exclusive (e.g. interested/not interested, larger/smaller/same size). When this was not possible and several answers could have been chosen, we replaced the multiple choice questions by a series of true/false statements, one per possible answer.

The full questionnaires are provided in appendix. Before looking at the results, some detail about how the survey was conducted and how the samples were defined needs also to be provided.

2.b. Survey information

The survey was made during the first semester of 2012. This survey is in fact composed by two distinct fieldworks. The first one was a face-to-face survey among the general population in the province of Liège (to the southeast of Belgium), which provided 667 questionnaires. These interviews were conducted by sociology students enrolled in an advanced course of survey methodology. The questionnaires with 75 questions were filled through 15-min interviews with random people. The sampling strategy was to use three types of quotas: gender, age (three age groups), education level (three education level). The fieldwork was strictly monitored in order to join the different population and to avoid omitting some special "hard-to-reach" respondents. Quotas were fairly well respected as our samples descriptions will show in the section below (Table 1).

The second fieldwork was a paper questionnaire self-completed by students from schools located in the French-speaking part of Belgium. This fieldwork provided us 2589 questionnaires filled by students from all levels. The questionnaires were filled during classes, without any possible help, in order to measure the actual level of knowledge of students. These questionnaires were a bit longer (115 questions) than the ones used for the public at large, enabling to analyze more aspects of science communication.

In this paper we will use terms "*public at large*" to qualify members of the sample that answer the face-to-face survey and "*students*" to qualify students that answered the paper questionnaire. The full sample, containing both public and students, will be described by "*respondents*" (used whenever there is no difference between the results of the two samples). It should be noted that the questionnaires remained anonymous.

2.c. Samples' description

**Public at large**

The first sample consists of 667 persons from the public at large. Quotas were applied to ensure that the distribution of gender, age category, and education level reproduced the known distribution of these parameters amongst the Belgian population (Table 1). This ensures that our analysis is as little biased as possible.

*Table 1: repartition of the main identification criteria for the public sample, and agreement with actual classification data. The sum of the percentages should be 100% but, for the observed values, the total may be less than 100% because some people do not know or do not wish to answer.*

| *Criteria* | *Classification* | *N (=667)* | *Observed repartition in the data* | *Actual repartition amongst Belgian population* |
|---|---|---|---|---|
| Gender | Male | 330 | 49.5 | 48.1 |
|  | Female | 337 | 50.5 | 51.9 |
| Age | 18-34 yr old (aka young) | 324 | 48.6 | 27.3 |
|  | 35-64 yr old (aka adult) | 210 | 31.5 | 52.9 |
|  | >65 yr old (aka retired) | 133 | 19.9 | 19.8 |
| Education | Primary school diploma | 49 | 7.3 | 10.4 |
|  | Secondary school diploma | 405 | 60.7 | 57.7 |
|  | Higher education diploma | 203 | 30.4 | 31.9 |

To classify the populations, several other criteria were available in our questionnaire: employment status, presence or absence of a scientific background, presence or absence of an astronomical background (astronomy courses, amateur astronomy activities), type of religion (incl. which supreme authority is recognized to explain the world) and assiduity of religious practices, degree of concern about ecology, and self-evaluation of cultural level.

**Students**

The second sample consists of 2589 students. Note that the teachers who conducted the survey were volunteers, i.e. this second sample is biased towards good-will science teachers, and they are often from generalist schools (i.e. not technical colleges): the results were thus expected to be slightly too optimistic. However, the results were found to be very similar to those from the public at large, which may be explained as astronomy not being taught in schools, so that we are very confident in these results. In any case, no sample is perfect, so that one should keep in mind that comparisons are always more robust than plain numbers.

*Table 2: Similar as Table 1, but for the student sample.*

| Criteria | Classification | N (=2589) | Observed repartition in the data |
|---|---|---|---|
| Gender | Male | 1285 | 49.6 |
|  | Female | 1287 | 49.7 |
| Type of school attended by student | Primary school | 61 | 2.4 |
|  | Secondary school | 1268 | 49.0 |
|  | Higher education | 1226 | 47.4 |
| Scientific formation | Yes | 1113 | 43.0 |
|  | No | 1347 | 52.0 |
| Faith | Non believer | 1108 | 45.8 |
|  | Catholic | 827 | 34.6 |
|  | Muslim | 118 | 5.2 |
|  | Other religion | 158 | 6.5 |

## 3. Results: appreciation of astronomy

3.a Interest

The first part of the questionnaire dealt with the appreciation of astronomy. The first question was about declaring an interest in astronomy: 5% of the public declared to be very interested in this subject, 26% quite interested, 46% slightly interested, and 23% not at all interested. The proportions are very similar amongst the students. One third of the respondents are thus interested, to some degree, in astronomy. This proportion is higher than the 20% measured in Eurostat barometers for the interest in astronomical or spatial discoveries, but the question is here more general, explaining the higher proportion.

The next step was to try to characterize and identify the interested people. We thus have examined how the proportions changed depending on the socio-demographic criteria mentioned in previous section. Results appear similar whatever the faith[1], level of diploma, or gender in the public sample. The same is found amongst students, except that the male students appear in a proportion 60% higher, compared to female students, in the very+quite interested category. The latter result is in line with previous studies since gender often appears as a strong discriminant for science matters (Eurostat barometers, Entradas et al. 2013) but the associated result for the public at large is especially surprising – one may wonder whether astronomy is different from other "hard science", but our study is not able to derive a definitive answer to this question.

Age seems to have only a small influence, though it is significant: the young people amongst the public appear less at the extreme answers (very/not at all interested) than the adults or retired people; the secondary school students appear less interested than other students. Those are usual sociological trends amongst the population: young adults have less strong opinions, and teenagers are less interested whatever the domain.

---
[1] Catholic vs non believer – note that the Muslims students appear slightly more at the extreme positions (interested/not at all interested) than Catholic or non-believer students, but this trend is not significant nor observed amongst the public at large.

As could be expected, the interest significantly increases with the (self-evaluated) cultural level, the fact of having a scientific background, of having followed a course in astronomy, or of being amateur astronomer. Maybe more surprising is the correlation of interest in astronomy with the ecological concern, but people with an open and curious mind are probably more interested than others, whatever the subject.

Going further, we want to know what is the exact subject that people are interested in. Since we had more time with the students, we could get into details with them. We first asked them whether they had heard about a list of astronomical subjects: nearly everyone had heard at least once about seasons, the Moon, stars, planets, or astronauts; at least two thirds of the students had heard about black holes, the origin of the Universe, or its content; half of them had heard about robots exploring space or the fate of the Universe, and only one third of them had heard about cosmology. The next step was to see whether these subjects actually interested them: apart from cosmology (which interested only one student in four), at least half of the students were interested in every proposed subject. The highest interest – subject judged interesting by 75% of students – is arisen by planets and stars, while the lowest interest (excluding cosmology) – about 50% of interested students– goes to seasons and space exploration, human or robotic. All areas of astronomy are thus considered interesting. Note that cosmology seems to suffer from its name: when "fate of the Universe" or "origin of the Universe" is used rather than the technical term of "cosmology", the percentages of interested students are much higher. This recalls that avoiding jargon, or technical terms, is always a golden rule in the outreach domain.

***In summary, one third of respondents are interested in astronomy. The typical profile of the most interested person is an adult with a science formation and an interest in other subjects (such as ecology). The interest in astronomy is broad, without much preference for a subdomain.***

3.b Information

A second set of questions are related to the search for information about astronomy. Half of the respondents answer that they already have searched such information by themselves. That means that some people that declare little or no interest in astronomy do try to gather information on the subject! The obvious follow-up is to see where they search for that information: we thus provided a list of media and asked whether they had used each one. Some results do not vary between the public and student samples: two-thirds of them have looked for information in specialized press (magazines, books) or TV, half of them in museums and exhibitions about astronomy[2], about one-third in the general press (magazines, books), and only 10% in radio programmes. The last three possibilities got slightly different results in the two samples: half of the public, but two-thirds of the students, searched information at schools; one third of the public, but half of the students, search for answers in the family/friend circle; and two thirds of the public, but 90% of the students, use internet to get answers. These results are of course not surprising, since young students are more likely to get first information from teachers or parents, and they are also more likely to use new media, such as internet.

---
[2] In fact, about half of respondents have already visited a planetarium, a museum or an exhibition linked to astronomy. In the Eurostat barometers of 2005, only 16% of people declared having visited a science museum and 8% a science exhibit but the visit had to take place in the last year, while our question was general, without time limits.

Trying to get published in the general press seems to be not a good investment: when people have a specific question, they prefer to search in specialized items (magazines or books or exhibitions) rather than in general ones. That does not mean that the generalized press is without interest for outreach professionals, but that it serves a different goal than giving access to information – it may be used, for example, to arise interest in an issue, leading people to search for info afterwards. The radio gets a very low score, which is probably due to the fact that it is not seen as a media to get scientific information – it is better suited for music or news – despite the fact that several radio programs dealing with science exist. Our results may be compared to the results of the 2001 Eurostat barometer on "Europeans, Science and Technology". At the time, the sources of science information, classified by decreasing importance were: TV, press, radio, school, scientific journals, internet. While we found similar results for TV, press and school, the situation has clearly changed today for radio and internet. This is indeed no surprise with the advent of on-line encyclopedias such as Wikipedia, as well as new technologies such as smartphones and tablets – internet is now part of everyday life, while this is less and less the case for traditional radio programs.

The appreciation of a subject is not only mirrored by a declared interest or by having searched for associated information. Encountering science can happen in more informal venues, such as in discussions with family and friends. One question in our survey therefore is whether the Universe is a good subject to discuss: only one third of the respondents discuss, at least from time to time, about cosmic issues whereas another third discuss of that subject very rarely and the last third never discuss about it. This can be compared to science (in general) as subject of discussion: the Eurostat barometer of 2005 indicates that about half of the public discusses about science and technology with friends, on a regular basis or occasionally. Astronomy is thus not the scientific subject first chosen in discussion – indeed, health, environment, and technology subjects scored higher in interest-wise questions. However, half of the respondents find the theme interesting to discuss, with an additional fourth being indifferent – only the last fourth consider the theme as little or not at all interesting. A large majority would thus not exclude discussing the subject, if someone else brings it on the table. In the same vein, when asked about the place of astronomy in teaching programmes, a strong majority of about half of the students would prefer to keep the same situation as today, but a significant one third would like to have an increased astronomy content in their courses. There is no gender or school level difference for this result.

*In summary, one person on two is willing to discuss about astronomy and has already searched for information – specific information are searched in specialized press or on the web.*

## 4. Knowledge and reasoning

4.a Scoring

The questionnaire contains a list of true/false statements related to basic astronomy. The details of these questions will be analyzed in the next subsection, but having such a list enable us to derive a simple "score"[3]. In this subsection, we want to take a detailed look at these global results. Before doing so, it is worthwhile to consider what people think they know. The results are strikingly different from the self-evaluation of the cultural level: while the median cultural level is 6 (out of 10, with 10 the maximum), the median astronomical level is only 3/10, i.e. 50% of the respondents evaluate their astronomical level to 3 or less. People therefore generally consider having a poor knowledge of astronomy. Turning to actual scores then yields a surprising result: people know more than they think! The average score is 7.2/10 for students and 7.4/10 for the public. This recalls the results of the 2005 Eurostat barometer on "Europeans, Science and Technology", which found a similar average score to a list of true/false statements on general science subjects: astronomy is thus no exception amongst sciences, and basic scientific knowledge is present amongst the public.

Globally, there is no striking difference between students and the public, except for the Moon influencing the tides, a fact known by 90% of the public, but only 78% of the students. The statements yielding the most correct answers are: "the Sun produces light", "star may die", and "stars only exist at night". The statements yielding the most incorrect answers are: "The Earth is closer to the Sun than to the Moon", "The moon produces light", "The Earth is closer to the Sun than to the stars", "Jupiter has its own moons".

*Figure 1: The percentage of correct answers for the list of true/false statements, for the public sample.*

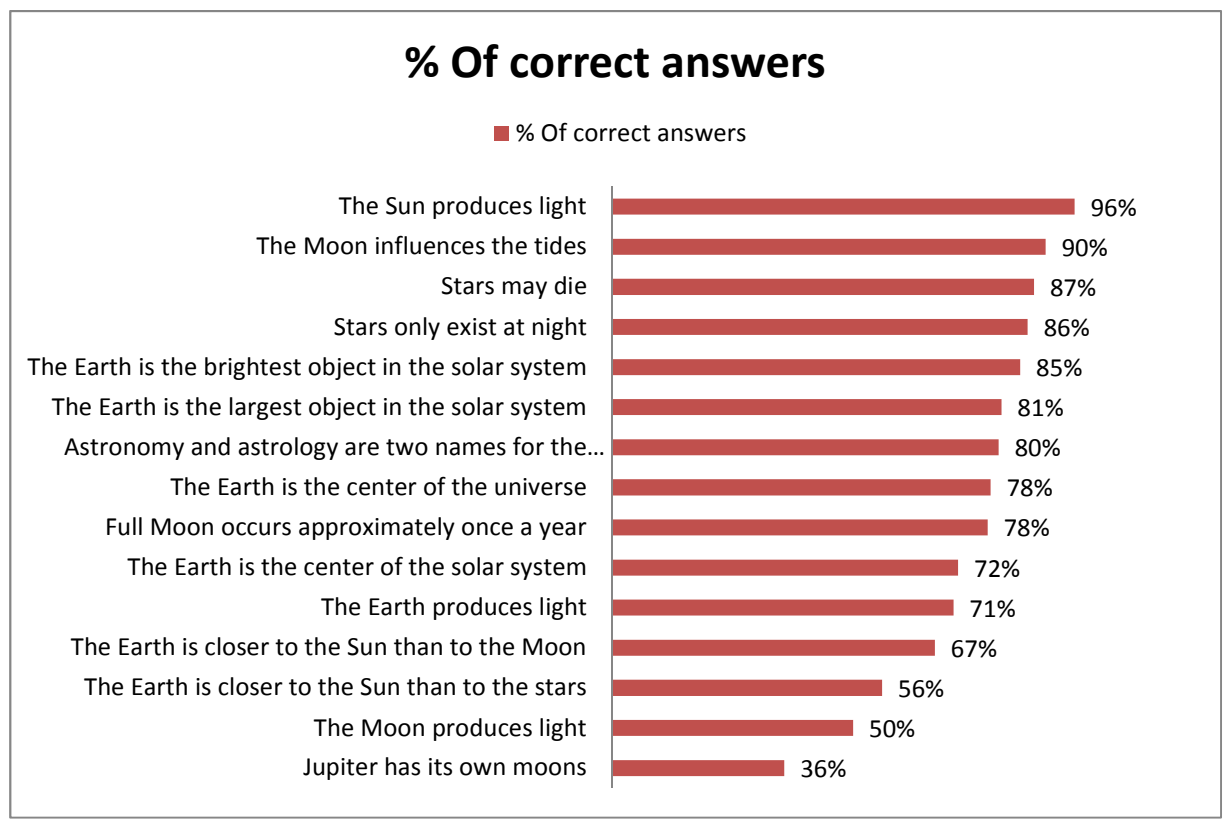

---
[3] Note that missing or "I don't know" answers were considered to yield zero points, as for an erroneous answer.

The scores depend on a number of factors. There is an obvious (and expected) positive correlation between the score and the self-evaluated cultural level, self-evaluated astronomical level, or interest in astronomy. Having a high level diploma (or being in a higher level school, for the students) and having a science background also help getting higher scores, as does the fact of having followed a course in astronomy – education is thus confirmed as a good way to learn things… As is also usual, women also score significantly less well than men. However, it should be noted that memory fails with time, since the score amongst the public also decreases with age (it should be noted that the questions were not linked to recent discoveries, all involved knowledge having been around for centuries: age is thus a priori a handicap). In short, the "best score" is obtained by a male student or worker with a university diploma.

*Figure 2: The average score for the public sample, as a function of several factors.*

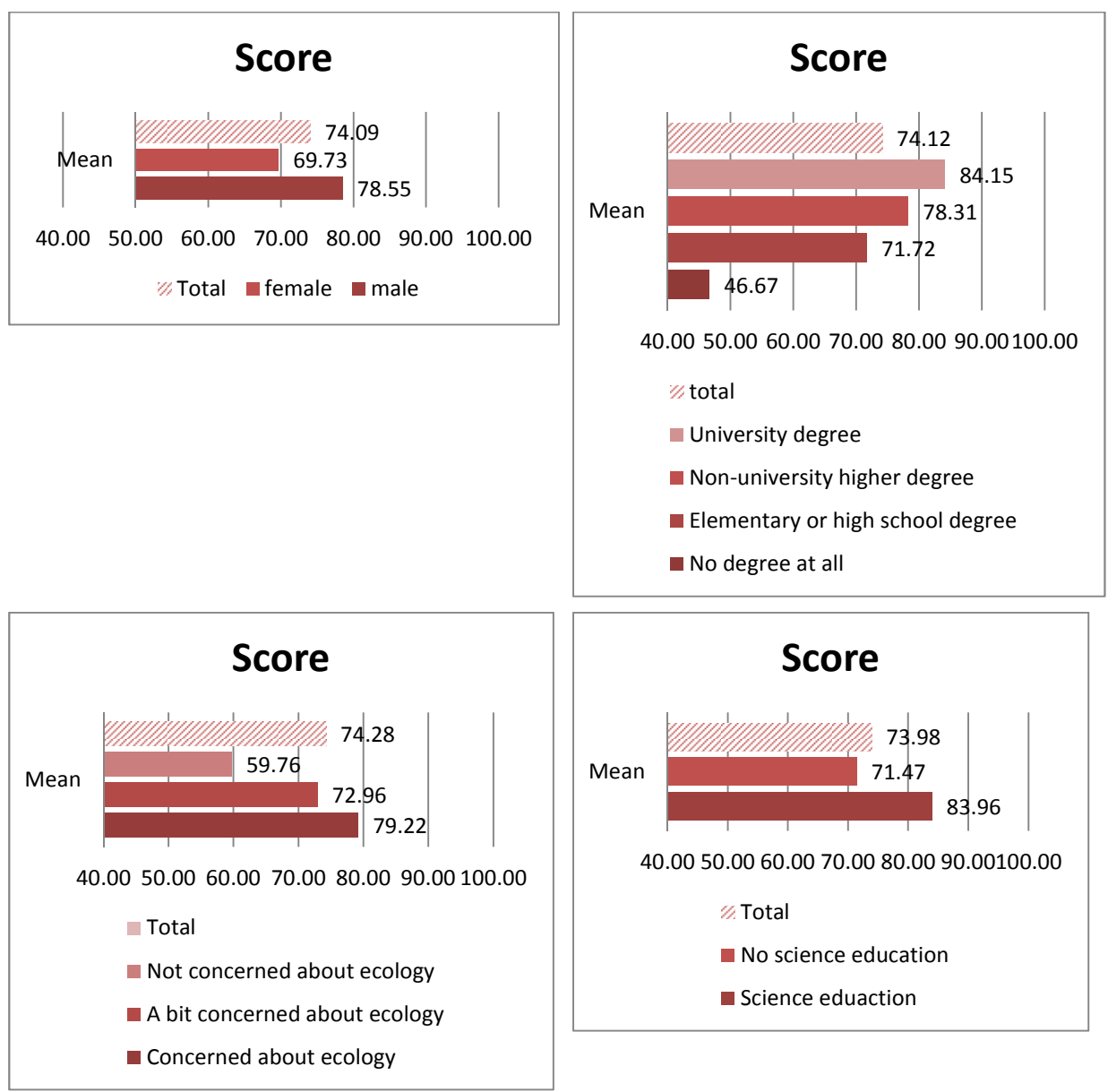

There are however three unexpected results. First, amateur astronomers give only slightly better answers than the non-amateur people. Second, the concern about ecology issues is again a discriminant factor, with higher scores for the more concerned people. Finally, faith appears as a discriminant factor too. For students, the average score is 7.2/10 for non-believers, 7.0/10 for Catholics, 6.2/10 for Muslims and 7.1/10 for followers of other religions. Similar results are found for the public sample. We have tried to find whether a hidden factor would be the cause of that change, but the additional criteria from the questionnaires yielded no answer. At first, it is a somewhat surprising result, especially for Muslims since several aspects of the Islamic ceremonials are linked to celestial phenomena. However, it is probable that other parameters should be taken into consideration, parameters not tested by our questionnaire. For example, Muslim students generally master less well the French language or how to access information; they also generally have parents with lower level diplomas compared to other students. These factors lead to a lower level of school success for these students, and may well play a role in our result.

4.b View of the cosmos

The questions were also aimed at studying the representations of the Universe by people. This is an important subject since preconceptions are the ground on which people build their knowledge. Knowing what is usually considered as big or small, close or distant, bright or faint, central or peripheral, important or not… represents a first step towards an efficient outreach. A conspicuous example in astronomy is of course whether people still live in the medieval "close universe", or have assimilated the consequences of "modern" observations.

Regarding the importance of Earth, the scientific revolution seems fully assimilated: the Earth is neither the center of the solar system, nor of the Universe for 70 to 80% of respondents; the Earth is not the biggest or brightest object in the solar system for a similar fraction. Moreover, the Earth is not producing light by itself for three quarters of the respondents. Getting into details with the students, it is indeed more difficult to put numbers: when asked for the typical size of Earth diameter, half of the students decline to give an answer. A quarter of them do yield acceptable answers, however: 11% give an estimate in the correct range, while another 10% mixes diameter and circumference, and another 5% diameter and radius. More surprising: while few consider the Earth to be small (only 3% estimate a diameter less than 5000km), about 10% see it very big (more than 50000km in diameter). Results are worse for the location of satellites: only 20% of students dare to answer to the question on the typical altitude of the International Space station (ISS): 2% see it below 100km (i.e. not in space, technically), 6% at a few hundreds km (the correct answer), 4% between 1000 and 10000km… and 9% above 10000km.

Of the two main objects of our sky, the Moon may be the most familiar, and maybe also the one most associated with phantasms. We thus wanted to see how people consider this celestial companion. The Moon is certainly known for its link with tides – 80 to 90% of respondents correctly associated both. A large majority (about 70%) also correctly indicate a Moon smaller than Earth, about 10% see it bigger and 7% of the same size. However, numbers are a problem, as usual. For the size of the Moon, compared to Earth, half of respondents do not answer, about 20% give an estimate in the correct range (2-5 times the Earth's), 12% a slightly larger ratio (5-10 times the Earth's) and about 5% indicate a very large difference between the two objects (with a Moon more than 100 times the Earth's size). For the distance of our satellite, 78% of the students decline to give a number, 3% see it at more than a million km, 6% provide a number below 10000km or between 10000 and 100000km, and 8% give the correct order of magnitude (a few hundreds thousands km), i.e., only one third of those who think they know give a correct answer. A last basic property of the Moon seems less understood: only half of the respondents correctly answered that the Moon does not produce light by itself. Finally, the possibility of other moons puzzles the respondents: about 40% do not know whether Jupiter has moons, and 20% think it does not.

*Figure 3: The size of the Moon and Sun, relative to Earth, in the public sample, with the correct range highlighted.*

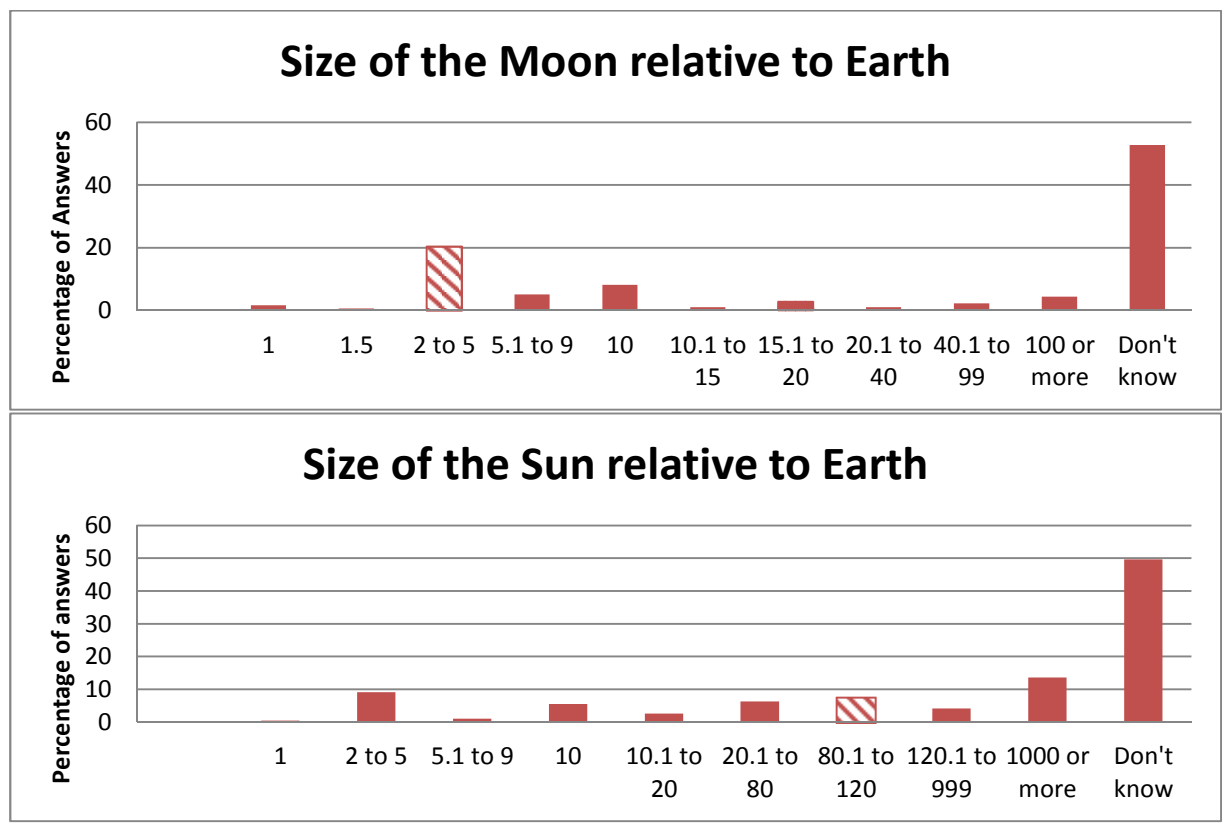

We have repeated this exercise for the Sun. It is an obvious source of light for more than 90% of respondents. It is also correctly seen as bigger than Earth for slightly less than 90% of respondents, and as more distant from Earth than the Moon for about two-thirds of the respondents. As usual, assigning orders of magnitude to these ideas is more difficult. When asked to precise their size estimate, half of the respondents decline answering, about 10% of respondents assign a size of a few times Earth's, about 15% a size of more than a thousand times Earth's, only 5-7% give the correct answer of 100 times Earth's. The solar size therefore seems to be more problematic than the Moon's. Distance to the Sun is even less known, since three quarters of students decline to answer, and there are similar fractions of them that assign a distance less than one million km, between one million and 10 millions km, of about 100 millions km, and more than a billion km: in short, they have no clue where the Sun is located.

Earth, Moon and Sun are usual terms, but planet, star or galaxy are not. Astronomers often use these words as if they were common and understood by all, but it remains to see whether this is the case. We begin with the first word: planets. As Earth, they are not seen as sources of light for about 70% of the students. However, their nature is not fully understood: about 10% of respondents do not know what they are, only 4% see them as similar to the Sun, and about half as similar to Earth… but one third see them as similar to the Moon. It is therefore not obvious for a large fraction of respondents that planets are "other earths". The confusion probably also extends to exoplanets, since "planets" is the problematic word. We therefore recommend always recalling what a planet is, just by a few words, to avoid misunderstandings.

Regarding stars, three-quarters of students consider them as able to produce their own light. Accordingly, their likeliness to the Sun (rather than Earth or Moon) is acknowledged by 60% of respondents. It is however significant that one person out of six do not know what stars are like or that the same proportion likens stars to the Moon. In addition, more than 80% of respondents correctly think that stars may die and that stars live not only during the nights. The basic ingredients of stellar knowledge are therefore present, with the word "star" apparently better understood than "planet". Assessing the distance to these objects is apparently more difficult. For half of respondents, stars are further from Earth than the Sun, but for a quarter of the respondents, stars are closer, and about 20% have no idea on the question. Asking a numerical estimate for the distance of the nearest star is even more difficult, with 5 out of six students declining to answer that question. Amongst those who answered, one third see the closest star at less than a million km and only one fifth provide the correct answer (4ly, or 4e13 km) or a value within an order of magnitude. The large distances of our cosmos are thus not imagined and, in a sense, we still live in a small, if not Earth-centered, Universe.

The last usual "jargon" word is galaxy. When asked to choose between solar system, cloud of gas or group of stars to define it, more than half of the students correctly choose the last item, the other two possibilities and the "no idea" receiving each about 15% of the votes. While this can be seen as a good step, one should then ponder what is meant by "group": 1, 5, 7, 10, 13, and 15% of the students consider that this stellar group has hundreds, thousands, millions, billions, hundreds billions, and millions billions components, respectively. In short, they have no idea of the size of a galaxy.

Despite not being jargon, the word astronomy itself may be subject to misunderstandings, and is often identified with astrology. One of the true/false statements was about this question. Surprisingly, 70 to 80% of respondents agree that astrology and astronomy are different. In fact, there are about twice more people that prefer to acknowledge their ignorance rather than stating that both words define the same thing. While this is encouraging, this is at odds with the results of the 2005 Eurostat barometer on "Europeans,

Science and Technology". In that European-wide study, 70% of people considered astronomy as scientific, but astrology reached a score of 41% (when replaced by "horoscope", the score dropped to 13%, showing that confusion still exists in people's mind between the two "astro". There is no reason which would explain a sudden change in recent years. Most probably the context of the questionnaire (e.g., choice of words) was sufficiently clear that the confusion was, this time, not made as often as usual.

Finally, another number is often used without restating it: the age of the Universe. A huge majority – five respondents out of six – correctly associate it with billions of years, rather than thousands or millions of years.

4.c Reasoning

Finally, a last set of questions was aimed at testing the actual understanding of astronomical concepts. Indeed, knowing roughly the meaning of a word does not necessarily imply understanding the actual concepts that this word subtended. Such analyses also help identifying misconceptions, with the possibility to correct them by good outreach practices before adding new blocks of knowledge. Three concepts were here tested: rotation of Earth on itself (day), rotation of Earth around the Sun (seasons), phases of the Moon.

It must be noted that we choose these three basic astronomical concepts because they are linked to everyday experience. Many questions could thus be easily answered by simple observation. As a first step, it is thus interesting to see whether people do observe or not, and how. Results of this survey are very encouraging: all respondents have already observed the sky with the naked eye, and half of them with binoculars. Maybe more surprising: one third of the public and half of the students already observed the sky through a telescope. This is most probably due to the success of cheap telescopes, sold as toys – indeed, this recent fashion is reflected by the fact that young people in the public are much more likely to have looked through a telescope than adults or retired persons. In what follows, we will call "observers" this group of respondents having used a telescope. Not surprisingly, these observers display a higher interest in celestial matters and higher scores. Quite naturally, this also has an impact on some observations: many people do not know that some planets can be seen by eye, or cannot recognize them in the sky; therefore, only one third of respondents declare having seen a planet at least once – but this percentage increases to 50% for the observers. Finally, one may wonder who these observers are: in fact, they generally are male (but only slightly more than female), of a high cultural level, in a science option or with a science formation, and concerned about ecology.

The rotation of Earth on itself is linked to several observable phenomena: alternation of day and night, existence of several time zones or of a "pole" star, changes in the position of the Sun throughout the day or with latitude. Questions linked to all these phenomena were asked, with relatively positive results though some misconceptions and lack of understanding are obvious. First, we investigate the day/night changes: about 60% of respondents correctly identify their origin (because Earth rotates on itself), about 20% of respondents misidentify day and year and answer that this alternation occurs because the Earth orbits around the Sun, and only 7% has a pre-Copernic view, believing that the Sun revolves around the Earth. Second, we investigated the world-wide changes. The concept of time zones is very well known with more than 80% of respondents knowing that at this moment, it is not the same hour in Australia or USA as in Belgium. However, the appearance of the sky seems more difficult to assess. When asked if, during the night, Australians would see the same stars as in Belgium, 40% of students correctly disagree, one third agrees, and a quarter does not know; the proportion of correct answers increases to about 50% when the situation in USA is considered. Of course, we do not expect many of the respondents to have actually personally experienced the American or Australian situation, but the questions are easy to answer in case of a true understanding of the Earth's rotation, geometry and geography: indeed, hourly and seasonal (see next paragraph) changes around the globe are apparently well known, but the consequences, on the stars, are not drawn. This is confirmed by a third set of questions, about the Pole star. A large majority (about 60%) of respondents know that it stays to the North, though a significant fraction of about 20% do not know. However, a similar majority see the Pole star as the brightest star of the sky, something that can be easily discarded if one has looked at the sky (only 16% of respondents provide a correct answer)… Astonishingly, the proportion of incorrect answers is larger for those who have looked through a telescope – telescope that needs to be aligned on the Pole star to work efficiently! Therefore, while everybody has looked at the sky, not many have truly *observed* it. Moreover, half of the respondents think that the Pole star can be seen from anywhere on Earth, showing that the actual concept of a "pole" star is not at all understood. Finally, we asked people about the Sun's position in our country. Half of the students and two thirds of the public have already seen the Sun towards the South (but about 15% has never seen the Sun in that "natural" direction for a northern hemisphere country such as Belgium). However, about 40% of respondents has already seen the Sun exactly up (i.e. at zenith) and about 20% towards the North, two locations which are not possible for the Sun as seen from Belgium. Again, the observers are slightly more likely to have observed these impossible configurations.

   We now turn to seasons and their cause. Before going into the heart of the subject, a first question relied on a simple observation, easy to gain by everyday experience: does the sun height above horizon change throughout the year? Half of the students and two-thirds of the public correctly affirm that the Sun appears higher in the sky during summer, only about 15-20% seeing no change in height. A second set of questions was related to world-wide changes in seasons: a large majority (about 70% of the students and 80% of the public) correctly realize that seasons are the same in USA and Belgium, or that they are different in Australia compared to Belgium. The last set of questions then directly investigate the cause of seasons: three possible causes were proposed (Sun height, length of day, distance to the Sun) with true/false answers along them. One third of students and half of the public correctly think that it is hotter in summer because the Sun is then higher in the sky, but about one third of the respondents disagrees with that statement. The proportions are similar for the proposal of heat linked to longer summer days. However, a shorter Earth-Sun distance is to blame for summer's heat for half of the respondents, with only 30% rejecting this possibility. This result, though a common error, is at odds with the large proportion of correct answers about seasons in USA or in Australia. Indeed, if the Sun-Earth distance was to blame, there would be no geographical variation in the seasons, no inversion between both hemispheres. In fact, it is true that the proportion of correct answers (the Sun-Earth distance is not to blame for the seasons) is higher in case of a correct answer to the questions about seasons in USA or Australia, but the proportion of incorrect answers is still much larger than 50%, even in case of a correct answer to the questions about oversee seasons. This apparent contradiction thus seems fully natural to most people. Trying to pin down this misconception, we also directly asked the students whether the Sun-Earth distance changes: it does for 41% of them, and does not for 24% of them. For the first group, a constant Sun-Earth distance would then imply no change in the seasons for 32% of them, the disappearance of seasons for 38%, and more strongly marked seasons for 14%. Another contradiction is seen here: half of those who affirm that the sun-Earth distance does not change points towards an Earth close to the Sun in summer as a cause for increased heat.

Finally, we examine more closely the Moon and its changes. As before, several questions linked to simple observations were asked. About 80% of respondents (even more amongst observers) have seen the Moon during the day – people are thus not influenced anymore by songs (like "Le Soleil a rendez-vous avec la Lune") or stories carrying the misconception of a day Sun and a night Moon that never meet. About 70% of respondents also correctly disagree to the statement that there is one full Moon per year. The detailed knowledge is, however, more surprising: while 42% of students correctly attribute a monthly timescale for changes in lunar shapes, 20% of them see the Moon change during the night – worse, that proportion increases for student observers. If changes are detected, the reason for the phases is however clearly unclear. About 30% of students simply did not answer, and the rest choose between four possible causes. The proposal that phases are linked to the Moon passing into Earth's shadow is chosen by about 40% of students assigning a nightly or monthly timescale for the phases. The (correct) proposal that phases are linked to the Moon orbiting the Earth is chosen by 28% of students assigning a nightly timescale and 38% of students preferring a monthly timescale. The alternative, a Moon orbiting the Sun as cause of the phases, is chosen by about 10% of the students. The possibility that phases are due to a dark side of the Moon is the least popular, with less than 5% of the votes. Students attributing a yearly timescale for the lunar phases spread rather uniformly between the possible causes of phases. Lunar eclipses are thus totally mixed with lunar phases, though simple observations readily show the Moon to be a thin crescent when close to the Sun, and full when opposite to it. Finally, a last observation was considered: the face of the Moon, which remains always the same. Again, despite being a simple observation, only one third of students (10% more amongst telescope observers) agree that the Moon show always the same face, with another third disagreeing, and the last third not knowing or refusing to answer. Those acknowledging the fact were asked for its possible consequence: half of them incorrectly infer that the Moon then does not rotate on itself, and only one third provide the opposite, and correct, answer.

## 5. Summary and conclusions

In 2012, we performed a detailed survey to assess the astronomical literacy level amongst French-speaking Belgian citizens. More than 3000 questionnaires were filled, with similar results in the two distinct samples (students, public at large). The aim was to analyze the interest and knowledge in astronomy, with the potential of improving outreach strategies.

Interest in astronomy is widespread, with one third of respondents declaring a (big or small) interest in this domain. Even half of respondents consider this domain as an interesting discussion subject and have visited a museum or exhibit related to astronomy. Note that the declared interest increases with the (self-evaluated) cultural level, ecological concern, or for respondents with a science background. To refine results, we confronted them to a list of subjects: overall, respondents show no strong preference and declare to find all of them interesting, with the highest scores for planets and stars and the lowest ones for robotic or human space exploration. Cosmology, black holes or astronauts, three very fashionable subjects in outreach activities, exert a priori no stronger appeal to the public or students: there is thus no need to focus *mostly* on these subjects, on the contrary, touching upon all astronomy subjects appears desirable.

Interested or not (!), half of respondents overall have searched at least once for information related to astronomy. The most efficient ways to reach them appear to be TV, specialized writings, internet, and astronomy-related museums or exhibitions. When focusing on the youngest ones, it is also useful to provide information to the teachers and the families.

To assess the overall astronomical literacy level, a list of statements was presented to respondents, with a true/false choice. This enabled us to calculate a "score" (one point being gained for each correct answer): the average score is about 7/10. The average score increases with the interest in astronomy, cultural level, ecological concern, and (achieved or current) school level. It is also higher for those with a science background, for those having attended an astronomy course. On the contrary, the average score is lower for older people or for women, and it is not significantly better for amateur astronomers. This may serve as a guide to adapt the outreach activities to the level to the audience.

The basic astronomical concepts (e.g. Earth not central, small Moon, big and shiny Sun, and galaxies as groups of stars) seem widely known, but details are often lacking. Numbers, for example, are always a problem: estimated sizes and distances, for those who dare giving a number, are rarely in the correct range. The sole exception is the age of the Universe, correctly identified to billions of years – the "young Earth" concept is thus not widespread in Belgium. Even common jargon is to be avoided, like e.g. "cosmology", simple words such as planet or star are worth re-defining quickly when doing astronomical outreach.

The basic astronomical facts (e.g. day/night change due to Earth's rotation, difference in hour and/or season in other countries) also appear widely known. However, knowing something does not necessarily mean understanding it. For example, the Pole star is thought to be seen from anywhere on Earth, or the Australian sky is often believed to be identical to that in Belgium. Of course, it is easier to recall a list of facts than gaining a deep understanding, able to correctly deduce all consequences of a given knowledge. However, a true understanding is a much better goal and outreach activities should thus aim at it rather than increasing the known fact list, which only provides superficial knowledge.

Being no exception, our respondents also display the usual misconceptions: changes in Sun-Earth distance as cause of seasons, Earth shadow as cause of lunar phases, Moon producing its own light, Moon not rotating because it always presents the same face to Earth. Interestingly, some of these deductions directly contradict answers to other questions, but that does not seem to be a problem. One of the best examples in that respect is the mismatch between summer heat due to changing Sun-Earth distance and opposite seasons in opposite hemispheres. This could be due to the tilted Earth image often used in teaching or outreach: the northern hemisphere is often said – or simply shown – to be "closer" to the Sun in winter. In any case, being aware of these misconceptions is crucial if one envisages an efficient outreach programme: indeed, it is well known that these deeply-rooted misconceptions must be debunked before anything correct can be built, or people will soon go back to their initial ideas.

Finally, astronomy is a purely observational science. Indeed, no direct experiment can be done on stars or galaxies, and one cannot reproduce a planet in the laboratory. It is therefore comforting that nearly everybody has looked at the night sky. Better still, half of them have done so through binoculars and up to half of the youngsters have used a telescope. However, the optimism must be moderate, since many people fail even very simple observations (Pole star is bright, Sun appears at zenith or North from Belgium, Moon shape changes during the night, Moon face is not always the same). In fact, looking at the sky does not imply *observing* it.

In summary, this survey shows that an interest for astronomy exists in Belgium, that people look at the sky, and that the basic concepts are already known. Some misconceptions are widespread, though, and should be taken into account for efficient outreach activities.

*Acknowledgments*

The authors acknowledge support from R. Cloots, Dean of the Science Faculty (ULg), and Pr M. Jacquemain (Sociology department, ULg). They are grateful to Mrs A. Finchi and C. Beck for encoding the answers in electronic format, to the sociology students for their help in conducting the survey of the public at large, and to volunteering teachers of many schools (H. Martinussen – Ecole d'Armurerie, CES Léon Mignon et Académie des Beaux-Arts, CES Léonard Defrance, Liège; C. Chevalier – IESPP Mons; R. Brisbois, head of Ecole Fond. Sart-Tilman; M. Deveux – St Andre Auvelais, C. Collette and P. Lafontaine – St Laurent Liège; T. Michetti – Institut Don Bosco Bruxelles; J.-F. Claeskens – AR Montegnée; S. Cicchirillo – Institut technique de Rance; M. Colemont and F. Monoyer – HEL section pédagogique; C. Boergoens – ICADI école de la Ville de Liège; F. Collette, head of A.R. Dour; O. Freches – A.R. Marchin; B. Surahy-Moureaux and A. Vandorselaer – Instituts St-Joseph et Sts-Pierre-&-Paul de Florennes; A. Baiverlin – A.R. Pepinster; G. Letawe and J. Remiche – Haute Ecole de la Province de Liège, catégorie paramédicale; H. Caps, P. Gerard, G. Hasbroeck, V. Pirenne, P. Poncin, and P. Deneye – University of Liège; C. Renotte, P. Gillis, and Y. Gossuin – University of Mons) for conducting the survey of their students.

# Appendix - Questionnaires

1. <u>Public at large:</u>

| 1) | Vous êtes de sexe …? | Masculin | 1 |
|---|---|---|---|
| | | Féminin | 2 |

| 2) | Année de naissance? | Ex : « 1978 » | |
|---|---|---|---|
| | | Refus *[ne pas lire]* | 9999 |

| 3) | Parmi les propositions suivantes, quel est le niveau du plus haut diplôme que vous avez obtenu ? | Sans diplôme | 1 |
|---|---|---|---|
| | | Primaire | 2 |
| | | Secondaire inférieur | 3 |
| | | Secondaire supérieur professionnel ou apprentissage | 4 |
| | | Secondaire supérieur technique, artistique | 5 |
| | | Secondaire supérieur général | 6 |
| | | Post secondaire non supérieur (formation de chef d'entreprise aux Classes moyennes, 7ème professionnelle …) | 7 |
| | | Supérieur non universitaire de type court | 8 |
| | | Supérieur non universitaire de type long | 9 |
| | | Supérieur universitaire | 10 |
| | | Doctorat avec thèse | 11 |
| | | Autre → Précisez | 12 |
| | | Ne sait pas *[ne pas lire]* | 88 |
| | | Refus *[ne pas lire]* | 99 |
| 4) Si « autre » à la question précédente, précisez | | | |

| 5) | Au cours de vos études (hors école secondaire), avez-vous suivi une formation en science exacte ? (chimie, physique, biologie, ingénieur, sciences de la vie, science de la santé, technicien scientifique,…) | Oui | 1 |
|---|---|---|---|
| | | Non | 2 |
| | | Ne sait pas *[ne pas lire]* | 88 |
| | | Refus *[ne pas lire]* | 99 |

| 6) | Si « oui » à la question précédente : Veuillez préciser en quelques mots le type de formation. | |
|---|---|---|

| 7) | Etes-vous …? | Non croyant | 1 |
|---|---|---|---|
| | | Catholique | 2 |
| | | Protestant | 3 |
| | | Orthodoxe | 4 |
| | | Juif | 5 |
| | | Musulman | 6 |

|   |   | Boudhiste | 7 |
|---|---|---|---|
|   |   | Hindouiste | 8 |
|   |   | autre religion chrétienne | 9 |
|   |   | autre religion orientale | 10 |
|   |   | Autre orientation religieuse | 11 |
|   |   | Ne sait pas *[ne pas lire]* | 88 |
|   |   | Refus *[ne pas lire]* | 99 |

| 8) | En dehors d'événements tels qu'un mariage, un baptême, un enterrement, une communion, combien de fois vous arrive-t-il d'avoir une pratique religieuse? | jamais | 1 |
|---|---|---|---|
|   |   | quelques fois par an | 2 |
|   |   | Au moins une fois par mois | 3 |
|   |   | Au moins une fois par semaine | 4 |
|   |   | Ne sait pas *[ne pas lire]* | 88 |
|   |   | Refus *[ne pas lire]* | 99 |

| 9) | Selon vous, qui est le plus légitime pour expliquer l'origine du monde ? | Une autorité religieuse | 1 |
|---|---|---|---|
|   |   | Le monde scientifique | 2 |
|   |   | Autre → Précisez | 3 |
|   |   | Ne sait pas *[ne pas lire]* | 88 |
|   |   | Refus *[ne pas lire]* | 99 |
| 10) | Si « autre » à la question précédente, précisez | | |

| 11) | Par rapport à l'ensemble de la population, vous considérez-vous comme quelqu'un de concerné par l'écologie ? | Tout à fait concerné | 1 |
|---|---|---|---|
|   |   | Un peu concerné | 2 |
|   |   | Pas du tout concerné | 3 |
|   |   | Ne sait pas *[ne pas lire]* | 88 |
|   |   | Refus *[ne pas lire]* | 99 |

| 12) | Sur une échelle de 1 à 10 et par rapport à l'ensemble de la population, à votre avis, quel est votre niveau de culture générale ? <br> *[Help : où 1 est le minimum et 10 est le maximum.]* | 1  2  3  4  5  6  7  8  9  10 | |
|---|---|---|---|
|   |   | Ne sait pas *[ne pas lire]* | 88 |
|   |   | Refus *[ne pas lire]* | 99 |

| 13) | Quelle situation correspond le mieux à votre état actuel? Choisissez une des propositions suivantes : | étudiant | 1 |
|---|---|---|---|
|   |   | Travailleur rémunéré | 2 |
|   |   | Pré-retraité, Retraité, pensionné | 3 |
|   |   | Personne au foyer (entretient le ménage et | ou s'occupe d'une personne dans le ménage: enfants, personnes âgées, ...) | 4 |
|   |   | Chômeur, demandeur d'emploi | 5 |
|   |   | En incapacité permanente | 6 |
|   |   | Autre → Précisez | 7 |
|   |   | Ne sait pas *[ne pas lire]* | 88 |

| | | Refus [ne pas lire] | 99 |
|---|---|---|---|
| 14) | Si « autre » à la question précédente, précisez | | |

| 15) | Etes-vous astronome amateur ? | Oui | 1 |
|---|---|---|---|
| | | Non | 2 |
| | | Ne sait pas [ne pas lire] | 88 |
| | | Refus [ne pas lire] | 99 |

| 16) | Avez-vous déjà suivi des cours spécifiques en astronomie (hors école secondaire) ? | Oui | 1 |
|---|---|---|---|
| | | Non | 2 |
| | | Ne sait pas [ne pas lire] | 88 |
| | | Refus [ne pas lire] | 99 |

| 17) | Etes-vous intéressé par l'astronomie ?<br><br>[Help : L'astronomie est l'étude du ciel, des étoiles et des planètes.] | Très intéressé | 1 |
|---|---|---|---|
| | | Plutôt intéressé | 2 |
| | | Peu intéressé | 3 |
| | | Pas du tout intéressé | 4 |
| | | Ne sait pas [ne pas lire] | 88 |
| | | Refus [ne pas lire] | 99 |

| 18) | Sur une échelle de 1 à 10 et par rapport à l'ensemble de la population, à votre avis, quel est votre niveau de connaissance en astronomie ?<br><br>[Help : où 1 est le minimum et 10 est le maximum.] | 1 2 3 4 5 6 7 8 9 10 | |
|---|---|---|---|
| | | Ne sait pas [ne pas lire] | 88 |
| | | Refus [ne pas lire] | 99 |

| 19) | Avez-vous déjà cherché des informations sur l'astronomie ? | Oui | 1 |
|---|---|---|---|
| | | Non | 2 |
| | | Ne sait pas [ne pas lire] | 88 |
| | | Refus [ne pas lire] | 99 |

| | Si oui à la question précédente, où avez-vous cherché des informations concernant l'astronomie ? | 20) | A l'école | **Oui (1)** | **Non (2)** | **S.O. (3)** |
|---|---|---|---|---|---|---|
| | | 21) | Auprès de ma famille ou de mes amis | **Oui (1)** | **Non (2)** | **S.O. (3)** |
| | | 22) | Sur internet | **Oui (1)** | **Non (2)** | **S.O. (3)** |
| | | 23) | Dans des livres ou des magazines spécialisés | **Oui (1)** | **Non (2)** | **S.O. (3)** |
| | | 24) | Dans des journaux et des magazines généralistes | **Oui (1)** | **Non (2)** | **S.O. (3)** |
| | | 25) | A la télévision | **Oui (1)** | **Non (2)** | **S.O. (3)** |
| | | 26) | A la radio | **Oui (1)** | **Non (2)** | **S.O. (3)** |
| | | 27) | Dans des musées ou des expositions | **Oui (1)** | **Non (2)** | **S.O. (3)** |
| | | 28) | Autre → précisez [ne pas lire] | **Oui (1)** | **Non (2)** | **S.O. (3)** |
| 29) | Si « autre » à la question précédente, précisez | | | | | |

| 30) | L'univers est-il un sujet de discussion fréquemment abordé entre vous et votre entourage ? | Très fréquemment abordé | 1 |
|---|---|---|---|
| | | Plutôt fréquemment abordé | 2 |
| | | Peu fréquemment abordé | 3 |
| | | Très peu abordé | 4 |
| | | Jamais abordé | 5 |
| | | Ne sait pas *[ne pas lire]* | 88 |
| | | Refus *[ne pas lire]* | 99 |

| 31) | A votre avis, le thème de l'univers est-il un sujet de discussion intéressant à aborder avec votre entourage? | Très intéressant | 1 |
|---|---|---|---|
| | | Plutôt intéressant | 2 |
| | | Indifférent | 3 |
| | | Peu intéressant | 4 |
| | | Très peu intéressant | 5 |
| | | Ne sait pas *[ne pas lire]* | 88 |
| | | Refus *[ne pas lire]* | 99 |

| | Voici une liste de pratique en lien avec l'astronomie, pour chacune d'entre-elles, pouvez-vous me dire si cela vous est déjà arrivé : | 32) | … Observer le ciel à l'œil nu | **Oui (1)** | **Non (2)** | **NSP (3)** |
|---|---|---|---|---|---|---|
| | | 33) | … Observer le ciel avec des jumelles | **Oui (1)** | **Non (2)** | **NSP (3)** |
| | | 34) | … Observer le ciel avec un télescope (ou une lunette) | **Oui (1)** | **Non (2)** | **NSP (3)** |
| | | 35) | … Visiter un musée ou une exposition d'astronomie ou un planétarium. | **Oui (1)** | **Non (2)** | **NSP (3)** |

| | Voici une liste de propositions, pour chacune d'entre-elles, pouvez-vous me dire si, à votre avis, elles sont vraies ou fausses : | 36) | La terre est plus proche du soleil que de la lune | **Vrai (1)** | **Faux (2)** | **NSP (3)** |
|---|---|---|---|---|---|---|
| | | 37) | La terre est plus proche du soleil que des étoiles | **Vrai (1)** | **Faux (2)** | **NSP (3)** |
| | | 38) | La Terre est le centre du système solaire | **Vrai (1)** | **Faux (2)** | **NSP (3)** |
| | | 39) | La Terre est le centre de l'Univers | **Vrai (1)** | **Faux (2)** | **NSP (3)** |
| | | 40) | Le Soleil produit de la lumière | **Vrai (1)** | **Faux (2)** | **NSP (3)** |
| | | 41) | La Lune produit de la lumière | **Vrai (1)** | **Faux (2)** | **NSP (3)** |
| | | 42) | La Terre produit de la lumière | **Vrai (1)** | **Faux (2)** | **NSP (3)** |
| | | 43) | La Lune influence les marées | **Vrai (1)** | **Faux (2)** | **NSP (3)** |
| | | 44) | Il y a environ une pleine Lune par an | **Vrai (1)** | **Faux (2)** | **NSP (3)** |
| | | 45) | L'astronomie et l'astrologie sont deux noms pour la même discipline | **Vrai (1)** | **Faux (2)** | **NSP (3)** |
| | | 46) | Jupiter possède ses propres Lunes | **Vrai (1)** | **Faux (2)** | **NSP (3)** |
| | | 47) | La Terre est l'objet le plus gros du système solaire | **Vrai (1)** | **Faux (2)** | **NSP (3)** |
| | | 48) | La Terre est l'objet le plus brillant du système solaire | **Vrai (1)** | **Faux (2)** | **NSP (3)** |
| | | 49) | Les étoiles peuvent mourir | **Vrai (1)** | **Faux (2)** | **NSP (3)** |
| | | 50) | Les étoiles n'existent que la nuit | **Vrai (1)** | **Faux (2)** | **NSP (3)** |

| 51) | A votre avis, par rapport à la Terre, le Soleil est ? | Plus grand | 1 |
|---|---|---|---|
| | | De même taille | 2 |
| | *[Help : On parle ici du diamètre de l'objet.]* | Plus petit | 3 |

| 51) (cont.) | | Ne sait pas *[ne pas lire]* | 88 |
|---|---|---|---|
| | | Refus *[ne pas lire]* | 99 |

| 52) | Si vous avez répondu « plus grand » ou « plus petit », pouvez-vos me dire combien de fois, à votre avis ? | **Fois** | |
|---|---|---|---|
| | | Ne sait pas *[ne pas lire]* | 88 |
| | | Refus *[ne pas lire]* | 99 |

| 53) | A votre avis, par rapport à la Terre, la Lune est ? | Plus grande | 1 |
|---|---|---|---|
| | | De même taille | 2 |
| | *[Help : On parle ici du diamètre de l'objet.]* | Plus petite | 3 |
| | | Ne sait pas *[ne pas lire]* | 88 |
| | | Refus *[ne pas lire]* | 99 |

| 54) | Si vous avez répondu « plus grande » ou « plus petite », pouvez-vos me dire combien de fois, à votre avis ? | **Fois** | |
|---|---|---|---|
| | | Ne sait pas *[ne pas lire]* | 88 |
| | | Refus *[ne pas lire]* | 99 |

| 55) | A votre avis, l'âge de l'univers se compte en … | En milliers d'années | 1 |
|---|---|---|---|
| | | En millions d'années | 2 |
| | | En milliards d'années | 3 |
| | | Ne sait pas | 88 |
| | | Refus | 99 |

| 56) | A votre avis, à quoi une planète ressemble-t-elle le plus ? | A la Lune | 1 |
|---|---|---|---|
| | | A la Terre | 2 |
| | | Au Soleil | 3 |
| | *[Help : Du point de vue de ses propriétés.]* | Ne sait pas | 88 |
| | | Refus | 99 |

| 57) | A votre avis, à quoi une étoile ressemble-t-elle le plus ? | A la Lune | 1 |
|---|---|---|---|
| | | A la Terre | 2 |
| | | Au Soleil | 3 |
| | *[Help : Du point de vue de ses propriétés.]* | Ne sait pas | 88 |
| | | Refus | 99 |

| | Selon vous, qu'est-ce qui rend l'étoile polaire particulière ? | 58) | … C'est la plus brillante | **Oui (1)** | **Non (2)** | **NSP (3)** |
|---|---|---|---|---|---|---|
| | | 59) | … On peut la voir depuis partout sur la Terre | **Oui (1)** | **Non (2)** | **NSP (3)** |
| | | 60) | … Elle reste au nord | **Oui (1)** | **Non (2)** | **NSP (3)** |

| 61) | Selon vous, à quel moment de l'année le Soleil est-il le plus haut dans le ciel ? | En été | 1 |
|---|---|---|---|
| | | En hiver | 2 |
| | | Toujours à la même hauteur quelle que soit la saison | 3 |
| | *[Help : Par rapport à l'horizon.]* | Ne sait pas | 88 |

|  |  |  | Refus | | | **99** |
|---|---|---|---|---|---|---|
|  | A votre avis, pourquoi fait-il plus chaud en été ? | **62)** | … Parce que la Terre est plus près du Soleil en été | **Oui (1)** | **Non (2)** | **NSP (3)** |
|  |  | **63)** | … Parce que le Soleil éclaire la Belgique plus longtemps en été | **Oui (1)** | **Non (2)** | **NSP (3)** |
|  |  | **64)** | … Parce que le Soleil est plus haut dans le ciel en été | **Oui (1)** | **Non (2)** | **NSP (3)** |

| **65)** | A votre avis, pourquoi y a-t-il une alternance entre le jour et la nuit ? | La Terre tourne sur elle-même | **1** |
|---|---|---|---|
|  |  | Le Soleil tourne autour de la Terre | **2** |
|  |  | La Terre tourne autour du Soleil | **3** |
|  |  | Les nuages bloquent régulièrement la lumière | **4** |
|  |  | Ne sait pas | **88** |
|  |  | Refus | **99** |

|  | A votre avis, à cet instant précis (à la minute où je vous parle), en Australie… | **66)** | C'est la même heure qu'ici | **Vrai (1)** | **Faux (2)** | **NSP (3)** |
|---|---|---|---|---|---|---|
|  |  | **67)** | C'est la même saison qu'ici | **Vrai (1)** | **Faux (2)** | **NSP (3)** |

|  | A votre avis, à cet instant précis (à la minute ou je vous parle), aux Etats-Unis .. | **68)** | C'est la même heure qu'ici | **Vrai (1)** | **Faux (2)** | **NSP (3)** |
|---|---|---|---|---|---|---|
|  |  | **69)** | C'est la même saison qu'ici | **Vrai (1)** | **Faux (2)** | **NSP (3)** |

|  | Voici une liste de propositions, pour chacune d'entre-elles, pouvez-vous me dire si vous les avez déjà constatées : | **70)** | En Belgique, avez-vous déjà vu le Soleil au zénith (c'est-à-dire juste au-dessus de votre tête) ? | **Oui (1)** | **Non (2)** | **NSP (3)** |
|---|---|---|---|---|---|---|
|  |  | **71)** | Depuis la Belgique, avez-vous déjà vu le Soleil au nord ? | **Oui (1)** | **Non (2)** | **NSP (3)** |
|  |  | **72)** | Depuis la Belgique, avez-vous déjà vu le Soleil au sud ? | **Oui (1)** | **Non (2)** | **NSP (3)** |
|  |  | **73)** | Depuis la Belgique, avez-vous déjà vu la Lune durant la journée ? | **Oui (1)** | **Non (2)** | **NSP (3)** |

| **74)** | Avez-vous des précisions à apporter à l'une de vos réponses ? | |
|---|---|---|

| **75)** | Avez-vous un commentaire à formuler sur ce questionnaire ? | |
|---|---|---|

## 2. Students:

| **76)** | Etes-vous | Un homme | **1** |
|---|---|---|---|
|  |  | Une femme | **2** |

| **77)** | Année de naissance? | Ex : « 2001 » | |
|---|---|---|---|
|  |  | Refus de répondre | **9999** |

| **78)** | En quel niveau ou cycle, êtes-vous ? | Primaire | **1** |
|---|---|---|---|
|  |  | Secondaire | **2** |

| | | Supérieur | 3 |
| | | Ne sait pas | 88 |
| | | Refus de répondre | 99 |

| 79) | En quelle année êtes-vous *(pour les étudiants en master, indiquer 4 pour master 1 et 5 pour master 2)* | 1 | 1 |
| --- | --- | --- | --- |
| | | 2 | 2 |
| | | 3 | 3 |
| | | 4 | 4 |
| | | 5 | 5 |
| | | 6 | 6 |
| | | Ne sait pas | 88 |
| | | Refus de répondre | 99 |

| 80) | Etes-vous dans une option scientifique ? *(ex. sciences fortes ; chimie, physique, biologie, ingénieur, sciences de la santé, technicien scientifique,…)* | Oui | 1 |
| --- | --- | --- | --- |
| | | Non | 2 |
| | | Ne sait pas | 88 |
| | | Refus de répondre | 99 |
| 81) | Si vous avez répondu « oui » à la question précédente : Veuillez préciser en quelques mots le type de formation. | | |

| 82) | Etes-vous …? | Non croyant | 1 |
| --- | --- | --- | --- |
| | | Catholique | 2 |
| | | Protestant | 3 |
| | | Orthodoxe | 4 |
| | | Juif | 5 |
| | | Musulman | 6 |
| | | Boudhiste | 7 |
| | | Hindouiste | 8 |
| | | autre religion chrétienne | 9 |
| | | autre religion orientale | 10 |
| | | Autre orientation religieuse | 11 |
| | | Ne sait pas | 88 |
| | | Refus de répondre | 99 |

| 83) | En dehors d'événements tels qu'un mariage, un baptême, un enterrement, une communion, combien de fois vous arrive-t-il d'avoir une pratique religieuse? | jamais | 1 |
| --- | --- | --- | --- |
| | | quelques fois par an | 2 |
| | | Au moins une fois par mois | 3 |
| | | Au moins une fois par semaine | 4 |
| | | Ne sait pas | 88 |
| | | Refus de répondre | 99 |

| 84) | Selon vous, qui est le plus légitime pour expliquer | Une autorité religieuse | 1 |
| --- | --- | --- | --- |
| | l'origine du monde ? | Le monde scientifique | 2 |
| | | Autre → Précisez | 3 |
| | | Ne sait pas | 88 |
| | | Refus de répondre | 99 |
| 85) | Si vous avez répondu « autre » à la question précédente, précisez | | |
| 86) | Par rapport à l'ensemble de la population, vous considérez-vous comme quelqu'un de concerné par l'écologie ? | Tout à fait concerné | 1 |
| | | Un peu concerné | 2 |
| | | Pas du tout concerné | 3 |
| | | Ne sait pas | 88 |
| | | Refus de répondre | 99 |
| 87) | Sur une échelle de 1 (minimum) à 10 (maximum) et par rapport à l'ensemble de la population, à votre avis, quel est votre niveau de culture générale ?[ | **1 2 3 4 5 6 7 8 9 10** | |
| | | Ne sait pas | 88 |
| | | Refus de répondre | 99 |
| 88) | Etes-vous astronome amateur ? | Oui | 1 |
| | | Non | 2 |
| | | Ne sait pas | 88 |
| | | Refus de répondre | 99 |
| 89) | Avez-vous déjà suivi des cours spécifiques en astronomie ? | Oui | 1 |
| | | Non | 2 |
| | | Ne sait pas | 88 |
| | | Refus de répondre | 99 |
| 90) | Etes-vous intéressé par l'astronomie ? | Très intéressé | 1 |
| | | Plutôt intéressé | 2 |
| | | Peu intéressé | 3 |
| | | Pas du tout intéressé | 4 |
| | | Ne sait pas | 88 |
| | | Refus de répondre | 99 |
| 91) | Sur une échelle de 1 (minimum) à 10 (maximum) et par rapport à l'ensemble de la population, à votre avis, quel est votre niveau de connaissance en astronomie ? | **1 2 3 4 5 6 7 8 9 10** | |
| | | Ne sait pas | 88 |
| | | Refus de répondre | 99 |
| 92) | Avez-vous déjà cherché des informations sur l'astronomie ? | Oui | 1 |
| | | Non | 2 |
| | | Ne sait pas | 88 |
| | | Refus de répondre | 99 |

| | Si vous avez répondu « oui » à la question précédente, où avez-vous cherché des informations concernant l'astronomie ? | 93) | A l'école | Oui (1) | Non (2) | S.O. (3) |
|---|---|---|---|---|---|---|
| | | 94) | Auprès de ma famille ou de mes amis | Oui (1) | Non (2) | S.O. (3) |
| | | 95) | Sur internet | Oui (1) | Non (2) | S.O. (3) |
| | | 96) | Dans des livres ou des magazines spécialisés | Oui (1) | Non (2) | S.O. (3) |
| | | 97) | Dans des journaux et des magazines généralistes | Oui (1) | Non (2) | S.O. (3) |
| | | 98) | A la télévision | Oui (1) | Non (2) | S.O. (3) |
| | | 99) | A la radio | Oui (1) | Non (2) | S.O. (3) |
| | | 100) | Dans des musées ou des expositions | Oui (1) | Non (2) | S.O. (3) |
| | | 101) | Autre → précisez | Oui (1) | Non (2) | S.O. (3) |
| 102) | Si vous avez répondu « autre » à la question précédente, précisez | | | | | |

| 103) | L'univers est-il un sujet de discussion fréquemment abordé entre vous et votre entourage ? | Très fréquemment abordé | 1 |
|---|---|---|---|
| | | Plutôt fréquemment abordé | 2 |
| | | Peu fréquemment abordé | 3 |
| | | Très peu abordé | 4 |
| | | Jamais abordé | 5 |
| | | Ne sait pas | 88 |
| | | Refus de répondre | 99 |

| 104) | A votre avis, le thème de l'univers est-il un sujet de discussion intéressant à aborder avec votre entourage? | Très intéressant | 1 |
|---|---|---|---|
| | | Plutôt intéressant | 2 |
| | | Indifférent | 3 |
| | | Peu intéressant | 4 |
| | | Très peu intéressant | 5 |
| | | Ne sait pas | 88 |
| | | Refus de répondre | 99 |

| | Voici une liste de pratique en lien avec l'astronomie, pour chacune d'entre-elles, pouvez-vous me dire si cela vous est déjà arrivé : | 105) | … Observer le ciel à l'œil nu | Oui (1) | Non (2) | NSP (3) |
|---|---|---|---|---|---|---|
| | | 106) | … Observer le ciel avec des jumelles | Oui (1) | Non (2) | NSP (3) |
| | | 107) | … Observer le ciel avec un télescope (ou une lunette) | Oui (1) | Non (2) | NSP (3) |
| | | 108) | … Visiter un musée ou une exposition d'astronomie ou un planétarium. | Oui (1) | Non (2) | NSP (3) |

| 109) | A votre avis, quelle place devrait avoir l'astronomie dans l'enseignement ? | Une place plus importante qu'actuellement | 1 |
|---|---|---|---|
| | | La même place qu'actuellement | 2 |
| | | Une place moins importante qu'actuellement | 3 |
| | | Ne sait pas | 88 |
| | | Refus de répondre | 99 |

| | Voici une liste de sujets, pouvez-vous indiquer desquels vous avez déjà | 110) | Les saisons | Oui (1) | Non (2) | NSP (3) |
|---|---|---|---|---|---|---|
| | | 111) | La Lune | Oui (1) | Non (2) | NSP (3) |
| | entendu parler | **112)** | Les planètes | **Oui (1)** | **Non (2)** | **NSP (3)** |
| | | **113)** | Les étoiles | **Oui (1)** | **Non (2)** | **NSP (3)** |
| | | **114)** | La cosmologie | **Oui (1)** | **Non (2)** | **NSP (3)** |
| | | **115)** | Les trous noirs | **Oui (1)** | **Non (2)** | **NSP (3)** |
| | | **116)** | L'exploration robotique de l'espace | **Oui (1)** | **Non (2)** | **NSP (3)** |
| | | **117)** | L'exploration humaine de l'espace (astronautes) | **Oui (1)** | **Non (2)** | **NSP (3)** |
| | | **118)** | Le destin de l'Univers | **Oui (1)** | **Non (2)** | **NSP (3)** |
| | | **119)** | L'origine de l'Univers | **Oui (1)** | **Non (2)** | **NSP (3)** |
| | | **120)** | Le contenu de l'Univers | **Oui (1)** | **Non (2)** | **NSP (3)** |
| | Voici une liste de sujets, pouvez-vous indiquer lesquels vous intéressent ? | **121)** | Les saisons | **Oui (1)** | **Non (2)** | **NSP (3)** |
| | | **122)** | La Lune | **Oui (1)** | **Non (2)** | **NSP (3)** |
| | | **123)** | Les planètes | **Oui (1)** | **Non (2)** | **NSP (3)** |
| | | **124)** | Les étoiles | **Oui (1)** | **Non (2)** | **NSP (3)** |
| | | **125)** | La cosmologie | **Oui (1)** | **Non (2)** | **NSP (3)** |
| | | **126)** | Les trous noirs | **Oui (1)** | **Non (2)** | **NSP (3)** |
| | | **127)** | L'exploration robotique de l'espace | **Oui (1)** | **Non (2)** | **NSP (3)** |
| | | **128)** | L'exploration humaine de l'espace (astronautes) | **Oui (1)** | **Non (2)** | **NSP (3)** |
| | | **129)** | Le destin de l'Univers | **Oui (1)** | **Non (2)** | **NSP (3)** |
| | | **130)** | L'origine de l'Univers | **Oui (1)** | **Non (2)** | **NSP (3)** |
| | | **131)** | Le contenu de l'Univers | **Oui (1)** | **Non (2)** | **NSP (3)** |
| | Voici une liste de propositions, pour chacune d'entre-elles, pouvez-vous indiquer si, à votre avis, elles sont vraies ou fausses : | **132)** | La Terre est plus proche du Soleil que de la Lune | **Vrai (1)** | **Faux (2)** | **NSP (3)** |
| | | **133)** | La Terre est plus proche du Soleil que des étoiles | **Vrai (1)** | **Faux (2)** | **NSP (3)** |
| | | **134)** | La Terre est le centre du système solaire | **Vrai (1)** | **Faux (2)** | **NSP (3)** |
| | | **135)** | La Terre est le centre de l'Univers | **Vrai (1)** | **Faux (2)** | **NSP (3)** |
| | | **136)** | Le Soleil produit de la lumière | **Vrai (1)** | **Faux (2)** | **NSP (3)** |
| | | **137)** | La Lune produit de la lumière | **Vrai (1)** | **Faux (2)** | **NSP (3)** |
| | | **138)** | La Terre produit de la lumière | **Vrai (1)** | **Faux (2)** | **NSP (3)** |
| | | **139)** | Les étoiles produisent de lumière | **Vrai (1)** | **Faux (2)** | **NSP (3)** |
| | | **140)** | Les planètes produisent de la lumière | **Vrai (1)** | **Faux (2)** | **NSP (3)** |
| | | **141)** | La Lune influence les marées | **Vrai (1)** | **Faux (2)** | **NSP (3)** |
| | | **142)** | Il y a environ une pleine Lune par an | **Vrai (1)** | **Faux (2)** | **NSP (3)** |
| | | **143)** | L'astronomie et l'astrologie sont deux noms pour la même discipline | **Vrai (1)** | **Faux (2)** | **NSP (3)** |
| | | **144)** | Jupiter possède ses propres Lunes | **Vrai (1)** | **Faux (2)** | **NSP (3)** |
| | | **145)** | La Terre est l'objet le plus gros du système solaire | **Vrai (1)** | **Faux (2)** | **NSP (3)** |
| | | **146)** | La Terre est l'objet le plus brillant du système solaire | **Vrai (1)** | **Faux (2)** | **NSP (3)** |
| | | **147)** | Les étoiles peuvent mourir | **Vrai (1)** | **Faux (2)** | **NSP (3)** |
| | | **148)** | Les étoiles n'existent que la nuit | **Vrai (1)** | **Faux (2)** | **NSP (3)** |

| 149) | A votre avis, par rapport à la Terre, le Soleil est ? | Plus grand | 1 |
|---|---|---|---|
| | | De même taille | 2 |
| | | Plus petit | 3 |
| | | Ne sait pas | 88 |
| | | Refus de répondre | 99 |
| 150) | Si vous avez répondu « plus grand » ou « plus petit », pouvez-vous préciser combien de fois, à votre avis ? <br>[ On parle ici du diamètre de l'objet.] | ………………………………………….. Fois | |
| | | Ne sait pas | 88 |
| | | Refus de répondre | 99 |

| 151) | A votre avis, par rapport à la Terre, la Lune est ? | Plus grande | 1 |
|---|---|---|---|
| | | De même taille | 2 |
| | | Plus petite | 3 |
| | | Ne sait pas | 88 |
| | | Refus de répondre | 99 |
| 152) | Si vous avez répondu « plus grande » ou « plus petite », pouvez-vous préciser combien de fois, à votre avis ? <br>[ On parle ici du diamètre de l'objet.] | ………………………………………….. Fois | |
| | | Ne sait pas | 88 |
| | | Refus de répondre | 99 |

| 153) | A votre avis, l'âge de l'univers se compte en … | En milliers d'années | 1 |
|---|---|---|---|
| | | En millions d'années | 2 |
| | | En milliards d'années | 3 |
| | | Ne sait pas | 88 |
| | | Refus de répondre | 99 |

| 154) | A votre avis, quelle est approximativement le diamètre de la Terre ? | ………………………………………….. km | |
|---|---|---|---|
| | | Ne sait pas | 88 |
| | | Refus de répondre | 99 |

| 155) | A votre avis, quelle distance sépare la station spatiale de la surface de la Terre ? | ………………………………………….. km | |
|---|---|---|---|
| | | Ne sait pas | 88 |
| | | Refus de répondre | 99 |

| 156) | A votre avis, quelle distance sépare la Terre du Soleil ? | ………………………………………….. km | |
|---|---|---|---|
| | | Ne sait pas | 88 |

| 157) | A votre avis, quelle distance sépare la Terre de la Lune ? | ………………………………………….. km | |
|---|---|---|---|
| | | Ne sait pas | 88 |

| 158) | A votre avis, quelle distance sépare la Terre de l'étoile la plus proche (hors système solaire)? | ………………………………………….. km | |
|---|---|---|---|
| | | Ne sait pas | 88 |

| 159) | A votre avis, à quoi une planète ressemble-t-elle le plus, du point de vue de ses propriétés (physiques) ? | A la Lune | 1 |
|---|---|---|---|
| | | A la Terre | 2 |
| | | Au Soleil | 3 |
| | | Ne sait pas | 88 |
| | | Refus de répondre | 99 |
| 160) | A votre avis, à quoi une étoile ressemble-t-elle le plus, du point de vue de ses propriétés (physiques) ? | A la Lune | 1 |
| | | A la Terre | 2 |
| | | Au Soleil | 3 |
| | | Ne sait pas | 88 |
| | | Refus de répondre | 99 |
| 161) | A votre avis, à quoi une galaxie ressemble-t-elle le plus ? | A un système solaire | 1 |
| | | A un nuage de gaz | 2 |
| | | A un groupe d'étoiles | 3 |
| | | Ne sait pas | 88 |
| | | Refus de répondre | 99 |
| 162) | Si vous avez répondu « groupe d'étoiles » à la question précédente, précisez combien d'étoiles composent ce groupe, selon vous : | Des centaines | 1 |
| | | Des milliers | 2 |
| | | Des millions | 3 |
| | | Des milliards | 4 |
| | | Des centaines de milliards | 5 |
| | | Des millions de milliards | 6 |
| | | Ne sait pas | 88 |
| | | Refus de répondre | 99 |

| | | | | | | |
|---|---|---|---|---|---|---|
| | Selon vous, qu'est-ce qui rend l'étoile polaire particulière ? | 163) | … C'est la plus brillante | **Oui (1)** | **Non (2)** | **NSP (3)** |
| | | 164) | … On peut la voir depuis partout sur la Terre | **Oui (1)** | **Non (2)** | **NSP (3)** |
| | | 165) | … Elle reste au nord | **Oui (1)** | **Non (2)** | **NSP (3)** |

| | | | |
|---|---|---|---|
| 166) | Selon vous, à quel moment de l'année le Soleil est-il le plus haut dans le ciel ?<br><br>*[On parle ici de hauteur par rapport à l'horizon.]* | En été | 1 |
| | | En hiver | 2 |
| | | Toujours à la même hauteur quelle que soit la saison | 3 |
| | | Ne sait pas | 88 |
| | | Refus de répondre | 99 |

| | | | | | | |
|---|---|---|---|---|---|---|
| | A votre avis, pourquoi fait-il plus chaud en été ? | 167) | … Parce que la Terre est plus près du Soleil en été | **Oui (1)** | **Non (2)** | **NSP (3)** |
| | | 168) | … Parce que le Soleil éclaire la Belgique plus longtemps en été | **Oui (1)** | **Non (2)** | **NSP (3)** |
| | | 169) | … Parce que le Soleil est plus haut dans le ciel en été | **Oui (1)** | **Non (2)** | **NSP (3)** |

| | | | |
|---|---|---|---|
| 170) | A votre avis, pourquoi y a-t-il une alternance entre le jour et la nuit ? | La Terre tourne sur elle-même | 1 |
| | | Le Soleil tourne autour de la Terre | 2 |
| | | La Terre tourne autour du Soleil | 3 |

| | | | | Les nuages bloquent régulièrement la lumière | **4** |
| --- | --- | --- | --- | --- | --- |
| | | | | Ne sait pas | **88** |
| | | | | Refus de répondre | **99** |

| | A votre avis, à cet instant précis (à la minute où vous lisez ces lignes), en Australie… | **171)** | C'est la même heure qu'ici | **Vrai (1)** | **Faux (2)** | **NSP (3)** |
| --- | --- | --- | --- | --- | --- | --- |
| | | **172)** | C'est la même saison qu'ici | **Vrai (1)** | **Faux (2)** | **NSP (3)** |
| | | **173)** | Le soir, on peut voir les mêmes étoiles | **Vrai (1)** | **Faux (2)** | **NSP (3)** |

| | A votre avis, à cet instant précis (à la minute où vous lisez ces lignes), aux Etats-Unis .. | **174)** | C'est la même heure qu'ici | **Vrai (1)** | **Faux (2)** | **NSP (3)** |
| --- | --- | --- | --- | --- | --- | --- |
| | | **175)** | C'est la même saison qu'ici | **Vrai (1)** | **Faux (2)** | **NSP (3)** |
| | | **176)** | Le soir, on peut voir les mêmes étoiles | **Vrai (1)** | **Faux (2)** | **NSP (3)** |

| | Voici une liste de propositions, pour chacune d'entre-elles, pouvez-vous indiquer si vous les avez déjà constatées : | **177)** | En Belgique, avez-vous déjà vu le Soleil au zénith (c'est-à-dire juste au-dessus de votre tête) ? | **Oui (1)** | **Non (2)** | **NSP (3)** |
| --- | --- | --- | --- | --- | --- | --- |
| | | **178)** | Depuis la Belgique, avez-vous déjà vu le Soleil au nord ? | **Oui (1)** | **Non (2)** | **NSP (3)** |
| | | **179)** | Depuis la Belgique, avez-vous déjà vu le Soleil au sud ? | **Oui (1)** | **Non (2)** | **NSP (3)** |
| | | **180)** | Depuis la Belgique, avez-vous déjà vu la Lune durant la journée ? | **Oui (1)** | **Non (2)** | **NSP (3)** |
| | | **181)** | Depuis la Belgique, avez-vous déjà vu une planète la nuit ? | **Oui (1)** | **Non (2)** | **NSP (3)** |

| **182)** | Depuis la Belgique, voit-on toujours la même face de la Lune ? | | | Oui | **1** |
| --- | --- | --- | --- | --- | --- |
| | | | | Non | **2** |
| | | | | Ne sait pas | **88** |
| | | | | Refus de répondre | **99** |

| | Si vous avez répondu « oui » à la question précédente alors … | **183)** | La Lune a toujours une face dans l'ombre | **Vrai (1)** | **Faux (2)** | **NSP (3)** |
| --- | --- | --- | --- | --- | --- | --- |
| | | **184)** | La Lune ne tourne pas sur elle-même | **Vrai (1)** | **Faux (2)** | **NSP (3)** |

| **185)** | A votre avis, la Lune change d'apparence *(ex. croissant, disque…)* | Au cours d'une nuit | **1** |
| --- | --- | --- | --- |
| | | Au cours d'un mois | **2** |
| | | Au cours d'une année | **3** |
| | | Ne sait pas | **88** |
| | | Refus de répondre | **99** |

| **186)** | Si vous avez répondu « au cours d'une nuit, d'un mois ou d'une année » à la question précédente, pourquoi la Lune change-t-elle d'apparence, selon vous ? | Parce que la Lune passe dans l'ombre de la Terre | **1** |
| --- | --- | --- | --- |
| | | Parce que la Lune a un côté noir | **2** |
| | | Parce que la Lune tourne autour du Soleil | **3** |
| | | Parce que la Lune tourne autour de la Terre | **4** |
| | | Ne sait pas | **88** |
| | | Refus de répondre | **99** |

| **187)** | La distance entre le Soleil et la Terre … | Change avec le temps | **1** |
| --- | --- | --- | --- |
| | | Est toujours la même | **2** |
| | | Ne sait pas | **88** |

|  |  | Refus de répondre | **99** |
|---|---|---|---|

| **188)** | Si vous avez répondu « change avec le temps » à la question précédente, alors si l'on imagine au contraire que le Soleil reste toujours à la même distance de la Terre … | Les saisons resteraient identiques | **1** |
|---|---|---|---|
| | | Il n'y aurait plus de saison | **2** |
| | | Les saisons seraient plus marquées | **3** |
| | | Ne sait pas | **88** |
| | | Refus de répondre | **99** |

| **189)** | Avez-vous des précisions à apporter à l'une de vos réponses ? | |
|---|---|---|

| **190)** | Avez-vous un commentaire à formuler sur ce questionnaire ? | |
|---|---|---|